\documentclass[iop]{emulateapj-rtx4}

\slugcomment{Accepted for publication in the Astrophysical Journal}
\lefthead{Rudnick et al.}
\righthead{Cluster Red Sequence Growth at $z<1.62$}

\usepackage{graphicx}
\usepackage{lscape}
\usepackage{natbib}

\newcommand\lstar{${\rm L}^\star$}

\newcommand{\ml}{${\rm M}/{\rm L}$}

\newcommand{\mlstarlamav}[1]{$\langle {\cal M_{\star}}/{\rm L}_{V}\rangle$}

\newcommand\msol{${\cal M_{\odot}}$}

\newcommand{\mstar}{${M^{\star}}$}

\newcommand\pz{$P(z)$}

\newcommand{\magrest}[1]{$#1_{rest}$}

\newcommand{\ltotrs}{$L_{tot,RS}$}

\def\gtsima{$\; \buildrel > \over \sim \;$}
\def\gsim{\lower.7ex\hbox{\gtsima}}
\def\ltsima{$\; \buildrel < \over \sim \;$}
\def\lsim{\lower.7ex\hbox{\ltsima}}

\renewcommand\micron{$\mu$m}

\newcommand{\spitzer}{\textit{Spitzer}}

\def\apj{ApJ}	
\def\aj{AJ}	
\def\mnras{MNRAS}	
\def\aap{A\&A}

\def\nat{Nature}

\def\apjl{ApJ}
	
\def\araa{ARA\&A}

\long\def\symbolfootnote[#1]#2{\begingroup%
\def\thefootnote{\fnsymbol{footnote}}\footnote[#1]{#2}\endgroup} 

\newcommand\clg{XMM-LSS~J02182-05102}

\begin{document}
\title{A Tale of Dwarfs and Giants: Using a $z=1.62$ Cluster to
  Understand How the Red Sequence Grew Over The Last 9.5 Billion
  Years.\thanks{Based on observations obtained at the
      European Southern Observatory using the ESO Very Large Telescope
      on Cerro Paranal through ESO program 386.A-0514(A).}}


\author{Gregory H. Rudnick\altaffilmark{2}, 
  Kim-Vy Tran\altaffilmark{3}, 
  Casey Papovich\altaffilmark{3},
  Ivelina Momcheva\altaffilmark{4,5}, 
  Christopher Willmer\altaffilmark{6}}

\altaffiltext{2}{The University of Kansas, Department of Physics and Astronomy, Malott room 1082, 1251 Wescoe Hall Drive, Lawrence, KS, 66045, USA; \texttt{grudnick@ku.edu}}
\altaffiltext{3}{George P. and Cynthia Woods Mitchell Institute for Fundamental Physics and Astronomy, and Department of Physics and Astronomy, Texas A\&M University, College Station, TX 77843-4242, USA}
\altaffiltext{4}{Observatories, Carnegie Institution of Washington, 813 Santa Barbara Street, Pasadena, CA 91101, USA}
\altaffiltext{5}{Currently at Department of Astronomy, 260 Whitney Avenue, Yale University, New Haven, CT 06511}
\altaffiltext{6}{Steward Observatory, University of Arizona, 933 N. Cherry Avenue, Tucson, AZ 85721, USA}

\begin{abstract}

We study the red sequence in a cluster of galaxies at $z=1.62$ and
follow its evolution over the intervening 9.5 Gyr to the present day.
Using deep $YJK_s$ imaging with the HAWK-I instrument on the VLT we
identify a tight red sequence and construct its rest-frame $i$-band
luminosity function (LF).  There is a marked deficit of faint red
galaxies in the cluster that causes a turnover in the LF.  We compare
the red sequence LF to that for clusters at $z<0.8$ correcting the
luminosities for passive evolution.  The shape of the cluster red
sequence LF does not evolve between $z=1.62$ and $z=0.6$ but at
$z<0.6$ the faint population builds up significantly.  Meanwhile,
between $z=1.62$ to 0.6 the inferred total light on the red sequence
grows by a factor of $\sim 2$ and the bright end of the LF becomes
more populated.  We construct a simple model for red sequence
evolution that grows the red sequence in total luminosity and matches
the constant LF shape at $z>0.6$.  In this model the cluster accretes
quenched blue galaxies from the field and subsequently allows them to
merge.  We find that 3--4 mergers among cluster galaxies during the
4~Gyr between $z=1.62$ and $z=0.6$ matches the observed luminosity
function evolution between the two redshifts.  The inferred merger
rate is consistent with other studies of this cluster.  Our result
supports the picture that galaxy merging during the major growth phase
of massive clusters is an important process in shaping the red
sequence population at all luminosities.

\end{abstract}
\section{Introduction}
\label{Sec:intro}

Understanding the formation of passive galaxies is an enduring problem
in astronomy.  These galaxies have very low star formation rates
(SFRs), little or no cold gas, and dominate the population of massive
galaxies in the local Universe \citep[e.g.][]{Kauffmann03}.  Their
colors are uniformly red and they lie in a distinct region of
color-magnitude (or color-mass) space called the red sequence.
Studies of local passive galaxies indicate that their stellar
populations are very old \citep[e.g.][]{Bower92b,Bower98}, with the
most massive passive galaxies having the oldest mean stellar ages
\citep[e.g][but see \citet{Trager08}]{Thomas05,Gallazzi06,Thomas10}.
A problem with these studies is that it is not immediately apparent
how to disentangle the ages of the stars from the time at which they
assembled into present day galaxies.  In a dramatic examples of the
pitfalls that are present, \citet{Lauer88} and \citet{Rines07} find
observational evidence for past and ongoing mergers of old stellar
systems in brightest cluster galaxies (BCGs).  This behavior is echoed
in the semi-analytic model of \citet{DeLucia07b} who show that BCGs
have old stellar ages but relatively recent epochs where the mass was
physically assembled.  More recently, however, observations have shown
that the observed stellar mass in BCGs has remained constant since
$z\sim1$ \citep{Whiley08,Collins09,Stott10,Stott11}, in contrast to
the model predictions and the implications from the merger remnants
seen in low redshift BCGs.  This highlights the difficulties inherent
in interpreting the evolution of massive galaxies using stricly
studies of the local universe.

Direct lookback studies of passive galaxies shed some light on their
origin and evolution.  For example, \citet{vanderwel05} use
fundamental plane observations of field galaxies to determine that
low-mass red galaxies at $z<1$ have younger mean stellar ages than
their more massive counterparts, similar to what is seen from the
local studies referred to above.  Efforts have also been made to
observe the buildup of the passive population in situ.  A population
of passive galaxies is seen as far back as $z\sim 2$
\citep{Labbe05,Daddi05,Cassata08,Kriek08,Brammer09}, indicating that
at least some passive galaxies already were in place by that time,
only 3.3~Gyr after the Big Bang.  Despite their early presence on the
stage, however, the number densities of passive galaxies evolved
dramatically at $z<2$ \citep{Labbe05,Kriek08} with a factor of $\sim
2$ in growth of number and mass densities at $z<1$ \citep{Bell04,
  Brown07, Faber07, Taylor09, Brammer11}.  A mass dependent growth is
nonetheless debated though, as some works have shown that the rate of
growth in the red galaxy population at $z<1$ is slower in more massive
galaxies \citep{Cimatti06,Brown07, Faber07} while \citet{Brammer11} do
not find a strong mass dependence in galaxies selected to have low
SFRs.  In support of some mass dependence to the growth of passive
galaxies, \citet{Brammer11} and \citet{Bundy06} find that there is a
stellar mass at which the passive and star-forming populations have
equal density and that this limit evolves to lower mass with the
passing of cosmic time.

The most favored explanation for this growth in the total number of
red sequence galaxies is via the transformation of star-forming
galaxies to passive ones, following a quenching episode
\citep{Blanton06,Bundy06,Faber07,Brammer11}.  In general, this
conclusion is consistent with the observed growth in number density on
the red sequence and the constant number density of blue star-forming
galaxies.  Unfortunately, however, the quenching mechanism has not
been conclusively identified and there are not enough bright blue
galaxies at $z<1.5$ to account for the observed evolution in the
massive red sequence population, if these blue galaxies were simply to
fade onto the red sequence \citep[e.g.][]{Bell04}.  One scenario to
explain the growth of massive red galaxies without massive
star-forming progenitors, is to grow them via the mergers of low-mass
galaxies.  These mergers have to be dissipationless (or ``dry'') in
order to preserve the red colors, isophote shapes, and low SFRs of the
massive galaxies \citep[e.g.][]{Bell04,Faber07}.  Such a dry merging
scenario has the added effect that it can grow the sizes of passive
galaxies, which may explain the large evolution in sizes implied by
some direct lookback studies
\citep{Daddi05b,McIntosh05,Trujillo06,vandokkum08,vanderwel08}.

From a theoretical perspective there are multiple candidates for
quenching, all of which involve either an active removal of cold gas
from a galaxy, or the prevention of gas cooling onto the galaxy.  Cold
gas may be removed violently during galaxy mergers via a combination
of feedback from supernovae, gravitational shocks, and that from an
active galactic nucleus \citep[e.g.][]{Cox04,Springel05}.  The supply
of cold gas coming from a hot gas halo may also be shut off by heating
from the central AGN \citep[e.g.][]{Croton06,McNamara07}.  While these
processes are promising, it is not yet clear how well AGN can
efficiently couple to the gas of the galaxy.  Alternatively,
environmental processes can either strip cold gas as a galaxy falls
into a hot intracluster or intragroup medium \citep[ram pressure
  stripping;][]{Gunn72} or strip the hot gas halo, thereby depriving
the galaxy of fuel for future star formation
\citep[strangulation;][]{Larson80}.  These mechanisms for
environmental quenching have been incorporated into cosmological
simulations by assuming that galaxies have their gas supply cutoff
once they enter a larger dark matter halo and become a ``subhalo'' or
``satellite'' galaxy.  The current (and simple) implementation of this
quenching has difficulty in matching the clustering and abundance of
red galaxies, implying that the model is too efficient at quenching
star formation in low-mass galaxies \citep[e.g.][]{Coil08}.

Clearly, an important way to constrain the ways in which star
formation can be quenched is to study in detail the buildup of the red
sequence as a function of galaxy mass or luminosity.  For example, it
is necessary to understand how the red sequence assembles in regimes
where different feedback modes may dominate, e.g. as a function of
environment.  Clusters of galaxies are one extreme of the dark matter
halo power spectrum and are promising testbeds for understanding the
mechanisms by which star formation can be suppressed.  An additional
benefit of studying clusters is that the trip to the red sequence in a
massive cluster is a one way street.  Cooling of gas from the ambient
intracluster medium is likely inefficient and the large relative
velocities in massive virialized systems make merging unlikely
\citep[but see][for examples of merging red galaxies in a merging
  cluster]{vandokkum99,Tran08}.  As we will discuss in this paper,
however, galaxy merging may actually have been an important player in
early growth phase of clusters.

Measurements of the luminosity function (LF) of red sequence galaxies
in clusters
\citep{DeLucia04,Tanaka05,Stott07,DeLucia07,Gilbank08,Rudnick09} and
the field \citep{Rudnick09} at $z<1$ have shown unambiguously that the
faint red sequence population builds up at later times than the bright
population.  Because light correlates well with mass for red sequence
galaxies the luminosity trend can also be interpreted as one with
stellar mass (although there are potential complications with this
simple \ml\ scaling due to the heirarchical growth of galaxies on the
red sequence \citep{Skelton11}).  This late build-up of the faint end
therefore implies that whatever quenched star formation may have done
so in low mass galaxies at preferentially later times.  By comparing
the total luminosities of clusters at $0.4<z<0.8$ with their likely
descendants at $z\sim 0$ from SDSS, \citet{Rudnick09} found that the
total light on the red sequence in clusters must have grown by a
factor of 1--3 over this span of time, similar to the inferred growth
of the field red sequence \citep[e.g.][]{Brown07}.

What is still unknown from an observational standpoint is how the red
sequence in clusters evolved at $z>0.8$.  Here we present a study of
the rest-frame $i$-band red sequence and its luminosity function down
to faint magnitudes in a $z=1.62$ cluster \clg\ (also known as
IRC0218).  This cluster was selected as an overdensity of galaxies
with \spitzer/IRAC colors indicative of being at high redshift
irrespective of their rest-frame colors \citep{Papovich08}.  It was
subsequently confirmed using spectroscopy and there are 11 galaxies at
$1.62<z_{spec}<1.63$ within 1 physical Mpc of the center
\citep{Papovich10,Tanaka10}.  Its spatial structure implies that it is
not relaxed and deep Chandra observations have found diffuse emission
indicative of a potentially underluminous intracluster medium \citep{Pierre11}.
This galaxy cluster has a strong red sequence at bright magnitudes
\citep{Papovich10} despite not being selected on the basis of red
rest-frame optical colors.  In addition it has a large abundance of
star-forming galaxies in the cluster core \citep{Tran10}.

In this paper we specifically explore the red sequence luminosity
function and its evolution.  In \S\ref{Sec:obs} we discuss the
observations, and the construction of a $K_s$-band selected catalog.
The color magnitude diagram is presented in \S\ref{Sec:CMD}.  We
discuss the future evolutionary path of \clg\ in \S\ref{Sec:mass_z}.
The red sequence LF is presented in \S\ref{Sec:LF} along with the
evolution of the LF shape and total luminosity of red galaxies in
clusters at $z<1.62$.  We discuss our results and the implications for
red sequence growth in \S\ref{Sec:discuss} and conclude in
\S\ref{Sec:summ}.

Throughout we assume ``concordance'' $\Lambda$-dominated cosmology
with
$\Omega_\mathrm{M}=0.3,~\Omega_{\Lambda}=0.7,~\mathrm{and~H_o}=70~{\rm
  h_{70}~km~s^{-1}~Mpc^{-1}}$ unless explicitly stated otherwise.  All
magnitudes are quoted in the AB system.

\section{Data and Observations}
\label{Sec:obs}

\subsection{Image Reduction and Calibration}

We imaged \clg\ in the $YJK_s$ bands using the HAWK-I instrument
\citep{Pirard04,Casali06} on the VLT in service mode in the fall of
2010 (PI: Tran; 386.A-0514(A)).  Our observations were split up into
three Observing Blocks (OBs) in $Y$, two in $J$, and one in $K_s$.
The total integration times were 9360s, 6240s, and 3000s, in $YJK_s$
respectively.  The image quality of our observations was excellent,
with FWHM=0\farcs52, 0\farcs60, and 0\farcs43 in $YJK_s$ respectively.

In order to avoid the gaps between the chips, the cluster was centered
in the middle of one of the chips and the field of view was rotated by
approximately 45 degrees to place other substructures on the centers
of different chips.  We dithered our observations in pseudo-random
pattern within a box 60\arcsec\ on a side.

\setcounter{footnote}{6}
The data were reduced using standard techniques for NIR imaging and
follow the steps taken for HAWK-I data from \citet{Lidman08} which
include dark subtraction, the application of twilight flats to remove
the pixel-to-pixel response, sky subtraction using the XDIMSUM package
in IRAF \footnote{IRAF is distributed by the National Optical
  Astronomy Observatories which are operated by the Association for
  Research in Astronomy, Inc., under the cooperative agreement with
  the National Science Foundation}, flux calibration, astrometric
calibration, and image combination.  SExtractor, SCAMP, and SWarp were
used for the astrometric calibration and image
combination \footnote{available at \texttt{http://terapix.iap.fr}}.

The images were photometrically calibrated using an 8\arcsec\ diameter
aperture on 2MASS stars within the field for $J$ and $K_s$.  The $Y$
band image was calibrated using a UKIRT standard observed on the same
night as our observations.  The photometric zeropoints have an
uncertainty of 0.02 mag.

\begin{figure*}[t]
\epsscale{1.15}
\plottwo{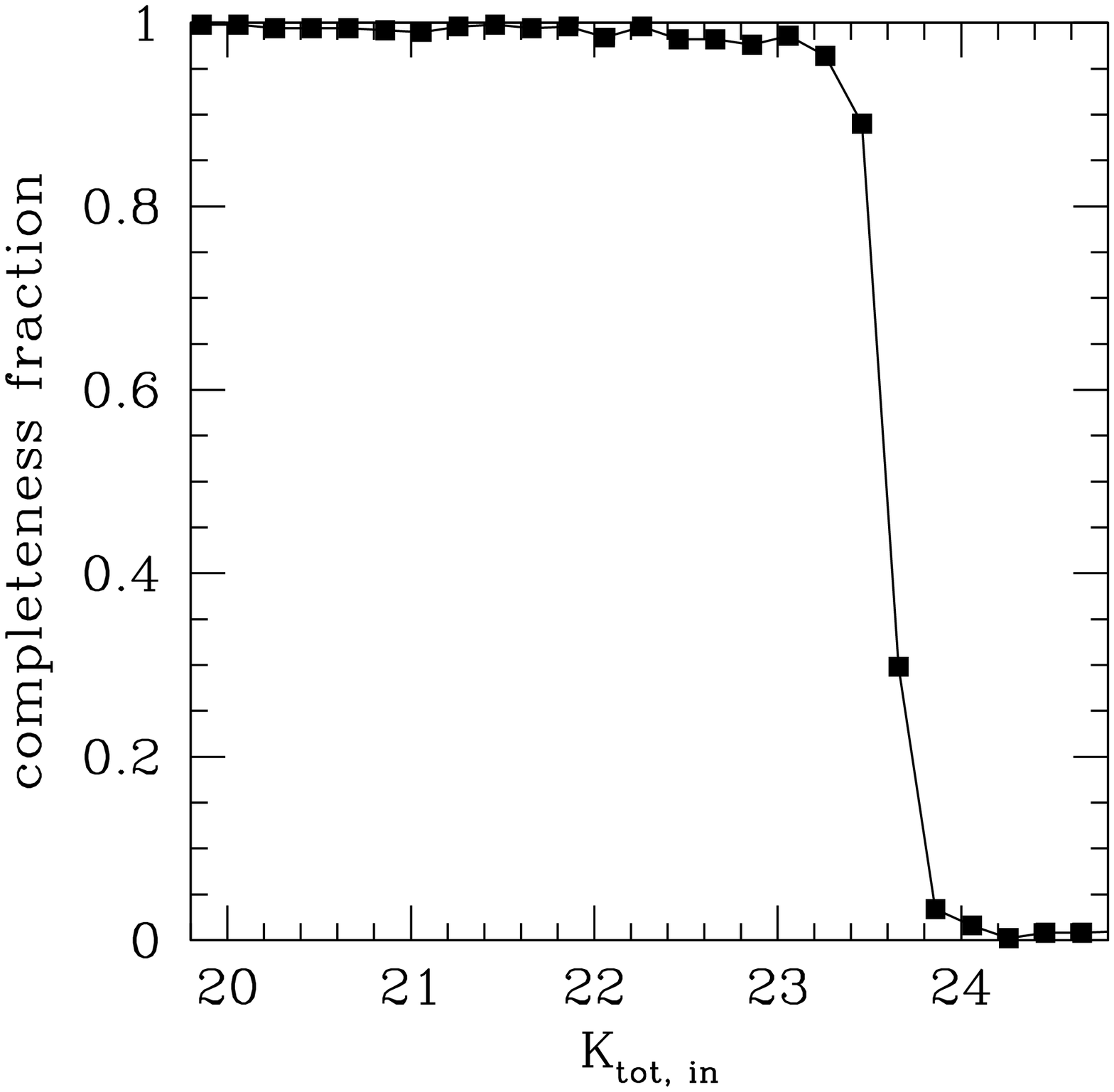}{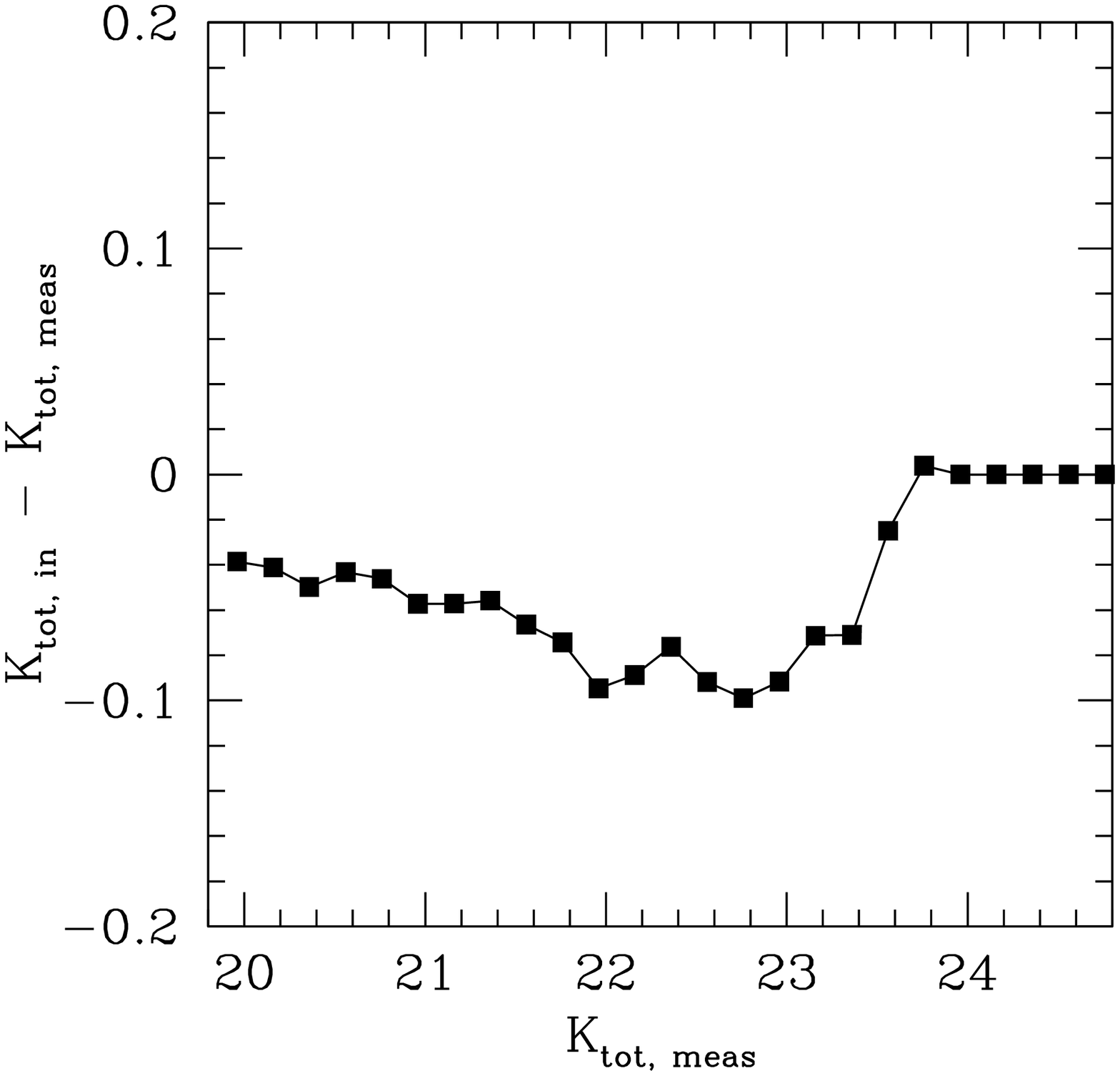}
\caption { The results of our point source completeness simulation.
\textit{Left panel:} The completeness fraction, defined as
  the ratio of detected sources over input sources, as a function of
  input magnitude. The completeness drops precipitously at $K_s>23.4$
  mag.    \textit{Right panel:} The error in total magnitude (including point
  source aperture correction) as a function of measured magnitude.
  The error in total magnitude for faint resolved sources is less than
  0.1 mag. }
\label{Fig:compl}
\end{figure*}

\subsection{Object Detection and Photometry}

For the purposes of measuring accurate colors, the $Y$ and $K_s$
images were convolved to the FWHM of the $J$ image, which had the
lowest quality.  This was done by matching the position of the stellar
locus for bright, uncrowded, and unsaturated stars between each of the
convolved $Y$ and $K_s$ bands and the $J$ band.

Objects were detected from the native seeing $K_s$ band image using
the SExtractor software (v2.5.0).  Before detection the $K_s$ image
was convolved within SExtractor with a Gaussian having a FWHM=5.0
pixels, or 0\farcs53, which corresponds to the PSF size.  This is
applicable for the faintest sources, which are likely unresolved.  We
experimented with different combinations of the detection threshold
(\texttt{DETECT\_THRESH}) and minimum number of required
interconnected pixels (\texttt{DETECT\_MINAREA}) and decided that the
parameters \texttt{DETECT\_THRESH}=2.0 and \texttt{DETECT\_MINAREA}=1
detected most obvious sources with no clear spurious detections. For
the purposes of this paper, spurious sources were identified as faint
$K_s$ detections with no counterpart in $Y$ or $J$.  We chose
\texttt{DETECT\_MINAREA}=1 because it can be translated
straightforwardly into a magnitude limit for point sources, which
simplifies understanding the detection in terms of a total magnitude
limit \citep{Labbe03}.  Admittedly, using such a low detection
threshold is somewhat ambitious but we wanted to make sure that we
went as deep as possible without incurring spurious counts.  We
therefore tested that most objects near our detection limit had
acceptable measurment uncertainties (see below) and that including the
faintest objects does not affect our conclusions in any way.  We
quantify our detection limit below using completeness simulations.

Object photometry was performed in ``dual-image'' mode, where sources
were detected on the unconvolved $K_s$ image and matched aperture
photometry was then performed on the seeing-matched images.
Following, e.g.  \citet{Labbe03}, we chose ``color'' apertures to
maximize the signal-to-noise ($S/N$) of our colors, while minimizing
systematic errors due to crowding.  For isolated sources we choose an
isophotal aperture unless the isophotal area is less than that of a
circle with $d=0.8$\arcsec.  That size corresponds to $1.4\times$FWHM,
which is the size that maximizes the S/N for a Gaussian profile.  In
the presence of crowding or blending we choose a
$d=1$\arcsec\ circular aperture so that the isophotal aperture size is
not corrupted.

Because of correlations in the sky measurements caused by sub-pixel
dithering, distortion corrections, and undetected background sources,
the measured pixel-to-pixel rms can underestimate the true uncertainty
in the flux of an object.  Since our objects are much fainter than the
sky, the uncertainty in the measured flux is dominated by the
uncertainty in the measured value of the sky.  Therefore, an
appropriate way to estimate the flux uncertainty is to measure the
uncertainty in the sky for each flux measurement.  We did this using
an empty aperture simulations described in e.g., \citet{Labbe03}.
Briefly, for a range of aperture radii we inserted 1000 randomly
placed non-overlapping circular apertures into each image, excluding
all objects and image boundaries.  The distribution of measured fluxes
for each empty aperture gives the real uncertainty in the sky
measurement.  As found by many authors, starting with \citet{Labbe03},
the measured noise scales super-linearly with aperture size and is
significantly larger than the expectation from pixel-to-pixel rms
assuming pure uncorrelated Gaussian fluctuations.  For each
object/band/aperture triplet we computed a linearized aperture size
and assigned the appropriate uncertainty as derived from the aperture
simulations.  Using these simulations, we measured a formal 5-sigma
limit for a $d=1$\arcsec\ aperture of 25.2, 24.8, and 24.1 mag in
$YJK_s$ respectively, accounting for a point source aperture
correction (see \S\ref{Sec:totmag}).

We require an object to have $>50\%$ of the effective exposure time to
be included in the analysis.  The gaps between the detectors is not
excluded by this cut and we therefore only exclude regions around the
edges of the image.  29\% of the sources in the original catalog were
excluded by this cut, but many of the sources with the lowest
effective exposure time are likely spurious.

Stars were rejected from the catalog using the SExtractor Stellarity
index.  We flagged as stars, objects with CLASS\_STAR$\geq$0.99 or
those objects with $J-K_S<0.95$.  This cut effectively removed stars
on the stellar locus in the $K_s$ vs. size plane.

\subsubsection{Total Magnitudes}
\label{Sec:totmag}

We measure total magnitudes in the $K_s$ image using the SExtractor
AUTO magnitude with a point-source aperture correction.  The AUTO
aperture scales with the first moment of the object radial flux
profile.  While a floor on the size of the AUTO aperture was set,
for small (and usually faint) objects it is nonetheless the case that
significant flux can be missed by the small AUTO aperture.  We compute
a minimal aperture correction for point sources using the methodology
described in \citet{Labbe03} and \citet{Rudnick09}.  

We found 7 bright isolated stars in our $K_s$ image and calculated
each of their curves of growth (COG), which were normalized at
$r=4$\arcsec, corresponding to the aperture used for our zeropoint
determination.  These were then averaged to obtain an average curve of
growth for stars in the image.  For each object the aperture
correction was computed from the COG using the circularized AUTO
aperture radius.  For our faintest objects ($23<K_s({\rm AUTO})<23.5$)
the mean aperture correction is 0.15 mag with an rms of 0.04 mag.
This point source correction can be regarded as the minimal correction
to an object's total flux. While this correction undoubtedly misses
some flux for extended objects it must be applied for all objects and
is likely appropriate for faint sources near the resolution limit.
Robust modeling of the aperture correction for the brighter and more
extended sources is difficult as we do not have an accurate measure of
their profile shape.  For the rest of the paper, all of our $K_s$
magnitudes include this aperture correction.

\subsubsection{Catalog Completeness}

\begin{figure*}
\epsscale{1.15}
\plottwo{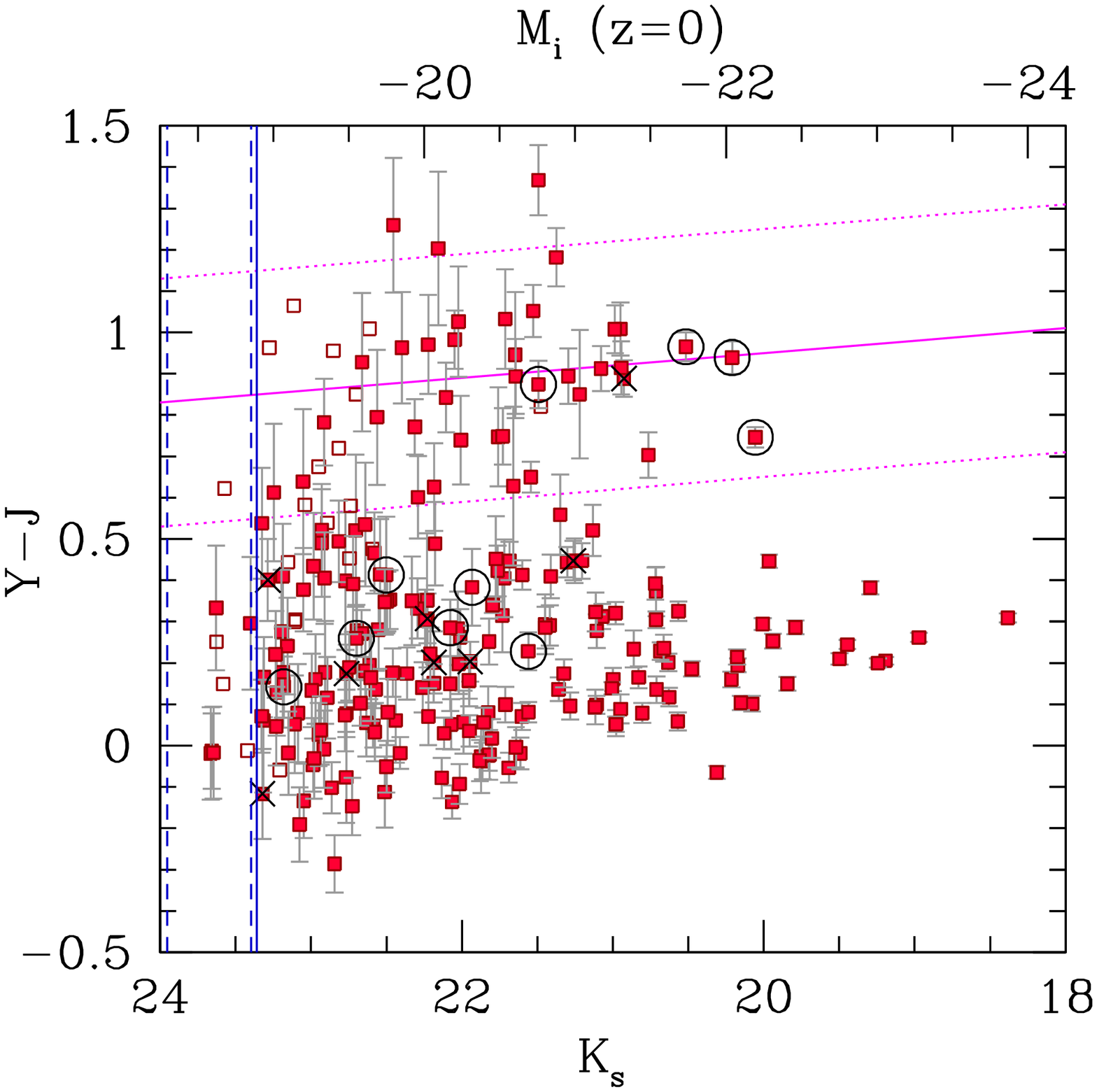}{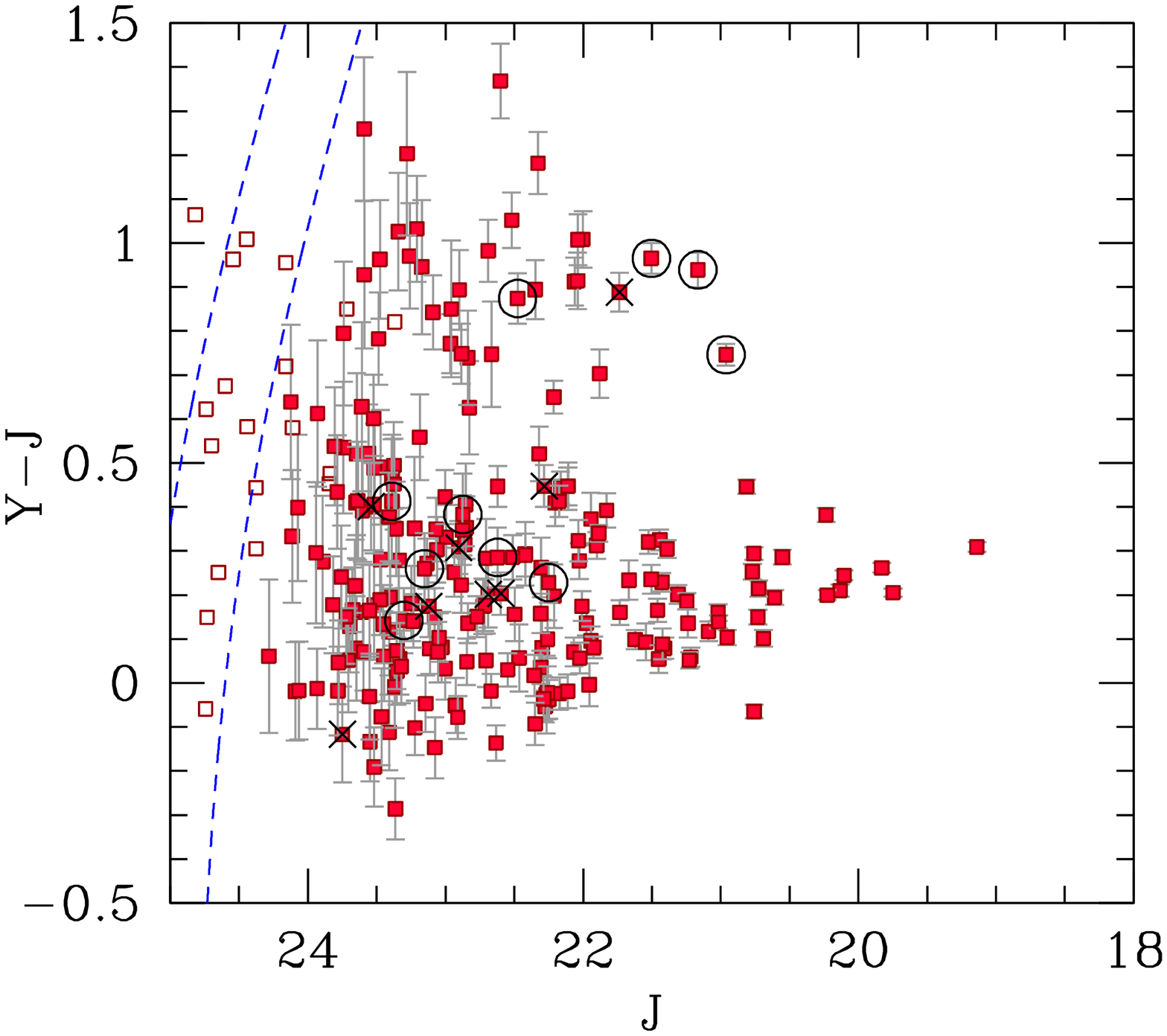}
\caption {Observed color magnitude diagrams for all of the objects
  within 2 arcminutes (1~Mpc) of the cluster center.  The catalog is
  selected in the $K_S$-band and the $J$ and $Y$-band data are 0.7 and
  1.0 mag deeper than $K_S$, respectively.  The open circles and
  crosses mark spectroscopically confirmed members and non-members
  respectively.  \textit{Left panel} The vertical blue solid line is
  the 90\% detection completeness limit and the vertical dashed blue
  lines correspond to the formal 5 and 3$\sigma$ limits in a
  $d=2$\arcsec\ circular aperture.  There is no unique relation
  between our $K_s$-band limit and a limit in $Y-J$ color but in both
  panels we use open squares to show those objects with a $Y-J$ color
  uncertainty greater than 0.2 magnitudes (see right panel).  The
  solid magenta line is a fit to all the galaxies with $Y-J>0.6$ and
  the dotted lines indicate $\pm 0.3$mag around this line.  The labels
  on the top of the figure indicate the absolute $i$-band magnitude
  derived from passively evolving the red sequence by 1.94 magnitudes
  from $z=1.62$ to $z=0$.  This can be directly compared with the
  magnitude in Figure~\ref{Fig:LF} and is only valid for red sequence
  galaxies.  \textit{Right panel} Here the curved dashed lines
  indicate the reddest colors for which we can achieve a 3 and 5-sigma
  measurement on the color in a $d=0.8$\arcsec\ circular aperture,
  which is appropriate for objects near our detection limit.  As is
  evident from these plots, the apparent lack of red objects at faint
  magnitudes is real and does not stem from either detection
  incompleteness nor from an inability to measure accurate colors.}
\label{Fig:CMD}
\end{figure*}

To assess our catalog completeness we performed a simulation to
determine how well we detect point sources, which should be good
analogs for the nearly unresolved objects near our detection limit.
We added 500 sources to the native seeing $K_s$ image, in batches of
100 to maintain the intrinsic crowding of the images, and detected the
objects with the same parameters used on the science data.  We then
measured the error in the magnitude as a function of measured
magnitude and the fraction of recovered sources as a function of input
magnitude.  The results of these simulations are shown in
Figure~\ref{Fig:compl}.  Our 90\% completeness limit is $K_s=23.4$ and
the amount of flux that we miss with our total magnitude estimate
(including aperture correction) is less than 0.1 mag for objects
brighter than this.  The 90\% completeness limit corresponds almost
exactly to the formal 5-sigma limit determined from our aperture
simulations.  It is also the magnitude at which the numbers of
observed galaxies starts to rapidly fall (see Figure~\ref{Fig:CMD}),
just as expected from our completeness simulation, where the
completeness falls from 90 to 30\% in just 0.2 mag.  This rapid
decline in completeness is a direct result of our aperture
corrections, as has been seen in other works \citep[e.g.][]{Labbe03,
  White05}.

\section{The Color Magnitude Diagram}
\label{Sec:CMD}

The $K_s$ vs. $Y-J$ color magnitude diagram for all galaxies within 2
arcminutes (1~Mpc) of the cluster center are show in the left panel of
Figure~\ref{Fig:CMD}.  At $z=1.62$ $Y-J$ corresponds to the age
sensitive $U-B$ rest-frame color as the $Y$ and $J$ filters straddle
the 4000\AA\ break.  The $K_s$ magnitude is very close to the
rest-frame $i$-band.  The plotted color-magnitude diagram (CMD)
includes the contribution of all galaxies along the line of sight.  In
constructing the LF we will statistically subtract the field to
measure the LF of the members but here discuss the full CMD.  Even
from this, however, there are several items worth commenting on.

There is a clear red sequence that is well separated from the blue
star-forming galaxies, as found in \citet{Papovich10} and
\citet{Tanaka10} using significantly shallower data.
\citet{Papovich10} demonstrated that the brightest red sequence
galaxies have rest-frame $U-B$ colors and a color scatter consistent
with a formation redshift of 2.35$\pm0.1$, indicating that they
experienced their last major episode of star formation $1.2\pm0.1$Gyr
before being observed.

It is possible that some of the galaxies on the red sequence have
colors dominated by dust extinction since some of them have
24\micron\ emission consistent with obscured star formation, or
possibly an AGN \citep{Tran10}.  Yet, $>80\%$ of the bright red
sequence galaxies have SED fits to their rest-frame $0.15{\rm \mu
  m}<\lambda<3.0{\rm \mu m}$ which do not indicate significant amounts
of dust extinction \citep{Tran10,Papovich11,Lotz11}.  Also, some of
the spectroscopic members are on the red sequence and a significant
fraction of these objects do not show evidence of emission lines in
their spectra \citep{Tanaka10}, although emission lines should have
been detectable at the cluster redshift \citep{Papovich10}.

We determine the slope and zeropoint of the red sequence by using a
robust line-fitting algorithm on all galaxies with $Y-J>0.6$ that have
errors in $Y-K$ and $K_s$ less than 0.2 mag. The slope and color at
$K_s=20$ are $-0.03\pm0.03$ and $0.95\pm0.13$.\footnote{The line fit
  remains unchanged if we include all galaxies with $Y-J>0.6$ and the
  resultant red sequence LF remains unchanged if we assume a constant
  color with magnitude for the red sequence.}

We find a striking lack of red galaxies at faint magnitudes.  This is
reminiscient of the lack of faint red galaxies in clusters at $z<1$
\citep{DeLucia04,Tanaka05,Stott07,DeLucia07,Gilbank08,Rudnick09} and
in a $z=1.46$ cluster \citep{Hilton09}, but now seen in a $z=1.62$
cluster.  It is worth addressing possible explanations for this
deficit.  This effect is not due to detection incompleteness as we
detect blue galaxies down to the 90\% completeness limit.  The
completeness could be a function of color if the faint red galaxies
have a larger size than blue galaxies at the same total magnitude but
we find this to be an unlikely possibility given the smaller size of
faint red galaxies at these redshifts compared to blue galaxies
\citep[e.g.][]{Zirm07,Toft07}.  

To assess the effect of color uncertainties on the perceived deficit
of faint red galaxies we plot the $Y-J$ vs. $J$ CMD in the right panel
of Figure~\ref{Fig:CMD}.  In this plot the curved lines represent the
reddest color to which we can measure the $Y-J$ color at the 3 and
5-sigma level in an aperture of a size appropriate for objects close
to our detection limit.  It is important to note that this line is not
a completeness limit but rather the limit to which colors can
accurately be determined.  The apparent lack of objects at magnitudes
slightly brighter than the 5-sigma line reflects our $K_S$-band
detection limit and the fact that our $J$ and $Y$-band data are 0.7
and 1.0 mag deeper than $K_S$, respectively.  The open symbols in both
panels are the same objects and correspond to those with color errors
in excess of 0.2 mag.  It is clear from this plot that the apparent
lack of objects at the faint end of the red sequence is neither an
issue of incompleteness nor does it result from large color errors, as
we could have measured colors accurately for objects in the empty
faint region of the red sequence if they were there.

It is important to realize that this deficit is robust to
considerations of cluster membership as we plot all galaxies within a
projected distance of 1~Mpc from the cluster.  Finally, it is possible
that galaxies with large color errors may be preferentially scattered
off the red sequence.  As we will discuss in \S\ref{Sec:LF}, when
performing the membership determination using statistical background
subtraction, assuming all of the galaxies with large errors lie on the
red sequence does not remove the presence of an observed deficit.
Therefore, the lack of observed faint red galaxies implies that these galaxies
are not present at any redshift.

\section{The Future Growth of \clg}
\label{Sec:mass_z}

\begin{figure}
\epsscale{1.2}
\plotone{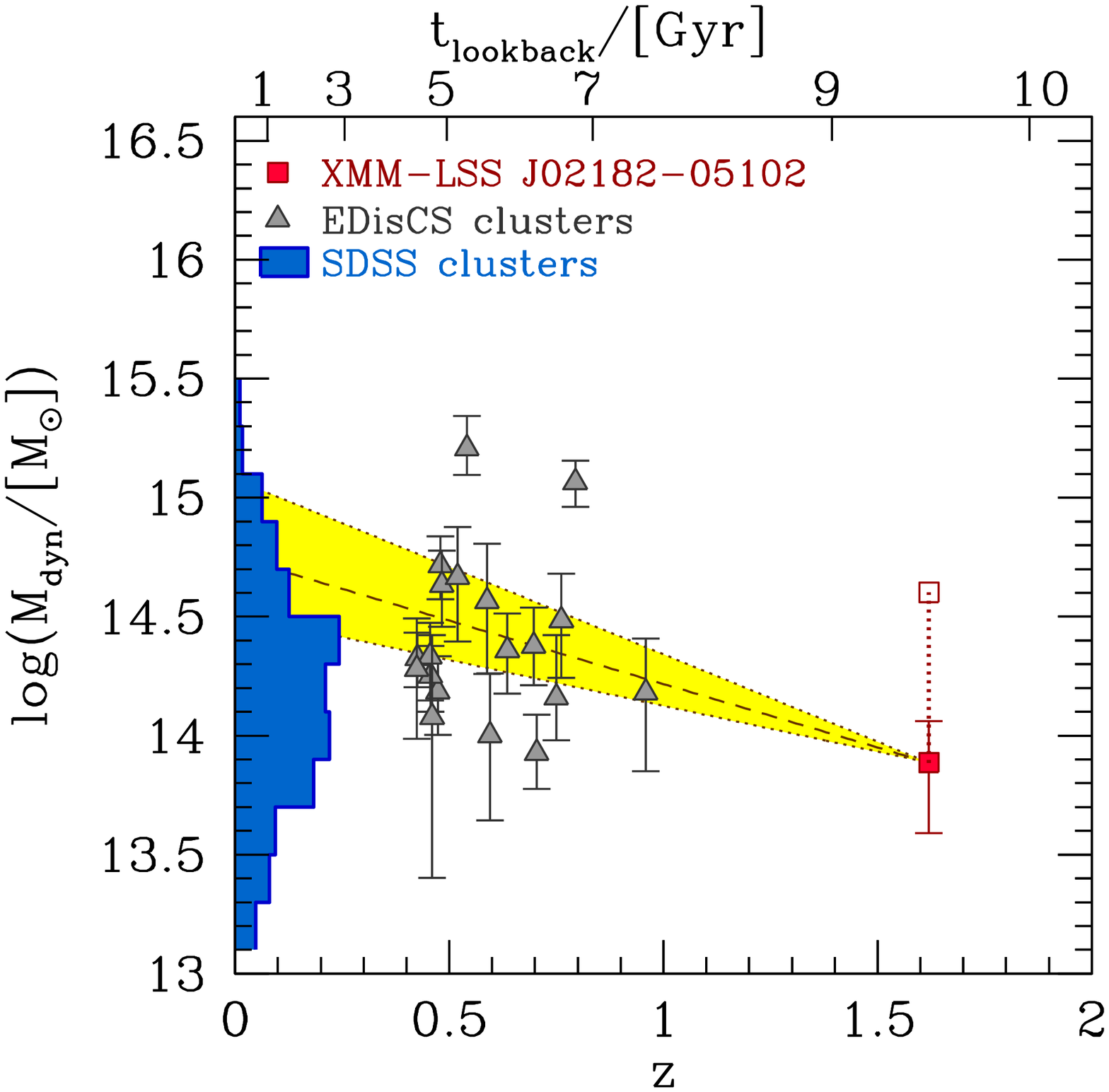}
\caption { The expected growth in mass of \clg\ to $z=0$.  The red
  solid square represents the x-ray derived mass for
  \clg\ \citep{Pierre11}.  The open square connected to the solid by a
  dotted line indicates the dynamical mass estimate from
  \citet{Papovich10}, which is significantly higher than the x-ray
  mass.  The disagreement between the two likely reflects the
  unrelaxed dynamical state of the system.  The gray triangles
  represent the EDisCS clusters.  The shaded histogram represents the
  distribution of dynamical masses from our local SDSS sample.  The
  diagonal dashed line gives the median expected growth in mass for
  \clg\ from the halo growth histories of \citet{Wechsler02} and using
  the halo concentrations of \citet{Bullock01}.  The dotted lines and
  yellow shaded region give the 68\% confidence limits on the growth
  given by the range in the halo concentrations for a cluster of this
  mass observed at $z=1.62$.  The likely descendants of \clg\ will be
  fairly typical clusters at all redshifts.}
\label{Fig:mass_z}
\end{figure}

To place \clg\ into an evolutionary context we must identify its
likely descendants at lower redshift.  To do this we must therefore
understand its expected growth in a hierarchically evolving Universe.
There are two mass measurements of \clg.  The first is derived from
the x-ray luminosity using the local Luminosity--Temperature and
Temperature--Mass relations, which yields $M\sim
7..7\pm3.8\times10^{13}M_\odot$\citep{Pierre11}\footnote{The x-ray
  luminosity for this cluster is not given in \citet{Pierre11}.}.  The
second is derived from the velocity dispersion of the galaxies and
$M\sim 4 \times10^{14}M_\odot$ \citep{Papovich10}.  As shown in the
appendix, both mass estimates are entirely consistent with that based
on the total red sequence light as calibrated from lower redshift
measurements.  Given the large degree of substructure, this cluster is
likely unrelaxed and we adopt the x-ray mass for the rest of the
paper.  We predict the mass growth of \clg\ using the halo growth
histories of \citet{Wechsler02}, which in turn uses the distribution
of halo concentrations from \citet{Bullock01} for a halo at this mass
and at this redshift.  \citet{Poggianti06} showed that these results
match the predictions from the Millennium simulation
\citep{Springel05b}.  In Figure~\ref{Fig:mass_z} we show this growth
and compare it to the masses of SDSS and EDisCS clusters as derived
from their galaxy velocity dispersions \citep{Milvang08}.  It is clear
that \clg\ will evolve into a typical (log$(M/M_\odot)\sim 14.3-15$)
system by $z<1$ even taking into account the dispersion in merger
histories of such massive halos.  We will use this projected growth in
subsequent sections when studying the evolution of the galaxies in
\clg.\footnote{Had we adopted the mass from the galaxy velocity
  dispersion, the predicted descendants of \clg\ would have been among
  the more massive clusters in the Universe at any redshift.  On one
  hand, such a massive nature for \clg\ may be unlikely given how rare
  such objects are.  Also, it is worth noting that the mass function
  of clusters at any epoch is very steep and that low-mass descendants
  will be preferred over high-mass ones.  Therefore, in addition to
  more precise mass measurements, a more careful treatment of the
  cluster mass function would be required to obtain a more accurate
  estimate of the typical descendant at each redshift.}

\section{The Red Sequence Luminosity Function}
\label{Sec:LF}

\subsection{Cluster Membership}
\label{Sec:member}
As very few spectroscopic redshifts are available in the core of the
cluster, and most of them are for blue galaxies, we identify which
galaxies are red sequence members using a statistical subtraction
technique.  \citet{Rudnick09} demonstrated that using
statistical background subtraction to isolate red sequence members in
clusters at $0.4<z<0.8$ yielded results that were identical to those
computed using accurate photometric redshifts.  An advantage of
statistical background subtraction for membership of faint galaxies is
that photometric redshift-based membership techniques often rely on
integrating the redshift probability distribution (\pz).  Faint
galaxies will have lower photometric $S/N$ and hence broader \pz\ and
the traditional method of establishing a threshold in the integrated
probability \citep{Brunner00} will more likely reject these object,
which is undesirable given the goals of this paper.  On the other
hand, statistical background subtraction does not have a formal
dependence on the $S/N$, other than being optimal when the
distribution of color and magnitude errors of the field and cluster
samples are identical, which is the case for our analysis.

Based on our experience \citep{Rudnick09} a suitable background
subtraction field needs to have comparable bandpasses (for red
sequence selection), depth, and total magnitude measurements to the
main survey field.  A wide area is also desireable.  We searched the
literature for data that satisfy these requirements but have not found
them.  For example, UKIDSS and GOODS while deep and wide, do not
include Y-band data at our depth.  We therefore decide to use the
outskirts of our HAWK-I image to define the background population.
While it is possible that the outskirts of our image may contain the
imprint of the associated large scale structure, it may indeed be
correct to subtract this "local" field rather than a field drawn from
a cosmic average.  

For the purposes of the subtraction, we define two regions, the
cluster at $r_{proj}<0.75$~Mpc\footnote{This is chosen to match the
  aperture used for the lower redshift comparison samples.  Its exact
  value does not affect our conclusions.}  and the field at
$r_{proj}>1.5$~Mpc.  After selecting galaxies within $\pm0.3$mag of
the red sequence, we bin the observed $K_s$ band magnitudes in both
regions and then subtract the field histogram from the cluster
histogram, normalizing the field histogram by the ratio of the field
area to the cluster area.  This results in an observed $K_s$ LF for
likely red sequence cluster members.

	\begin{deluxetable}{ccccc}
\tablecaption{Rest-frame $i$-band Luminosity Function of \clg}
\tablewidth{0pt}
\tablehead{\colhead{$M_{i,bright}$} & \colhead{$M_{i,faint}$} & \colhead{$\Phi$} & \colhead{$\delta \Phi_-$} & \colhead{$\delta \Phi_+$}}
\startdata
        -24.5   &    -24.0  &     1.0   &   0.83   &    2.41\\
        -24.0   &    -23.5  &    0.85   &   0.85   &    2.36\\
        -23.5   &    -23.0  &     5.08  &     2.17 &      3.47\\
        -23.0   &    -22.5  &     5.85  &     2.34 &      3.62\\
        -22.5   &    -22.0  &     2.79  &     1.56 &      2.94\\
        -22.0   &    -21.5  &     0     &      0   &        2.0\\
        -21.5   &    -21.0  &     2.19  &     1.36 &      2.78\\
        -21.0   &    -20.5  &    0.54   &   0.54   &    2.24
\enddata
\label{LF_tab}
\tablecomments {$\Phi$ corresponds to the number of galaxies in the
  magnitude range specified with no evolution corrections applied.
  The non-integer values are a byproduct of our statistical background
  subtraction technique.  The last two columns are the positive and
  negative uncertainty in this number as determined from Poisson
  uncertainty on the counts.  There is one magnitude bin with no
  galaxies detected.}
	\end{deluxetable}

\subsection{The Luminosity Function}
\label{SubSec:LF}

We construct the LF in the rest-frame $i$-band for comparison with
lower redshift studies \citep[e.g.][]{Rudnick09}.  To do this we take
advantage of the fact that the $K_s$ band is very close to the
redshifted rest-frame $i$-band filter.  Using the technique described
in \citet{Rudnick03} to compute \magrest{i} we find that the
k-corrections for red sequence galaxies at the cluster redshift are
consistent with a constant $m_{K_s}-M_{i}=44.13$ with a 0.03 mag
dispersion.  We apply this correction to the observed LF to obtain a
rest-frame \magrest{i}-band LF for red sequence cluster members.  

The shape of this LF is insensitive to the exact slope of the red
sequence and the width of our red sequence cut.  We also test how the
LF would change if we assumed that all galaxies with $Y-J$ errors more
than 0.2 mag were all on the red sequence.  Since this adds similar
numbers of galaxies to the field and cluster CMDs the shape of the
subtracted cluster LF is unchanged.

\begin{figure}
\epsscale{1.20}
 \plotone{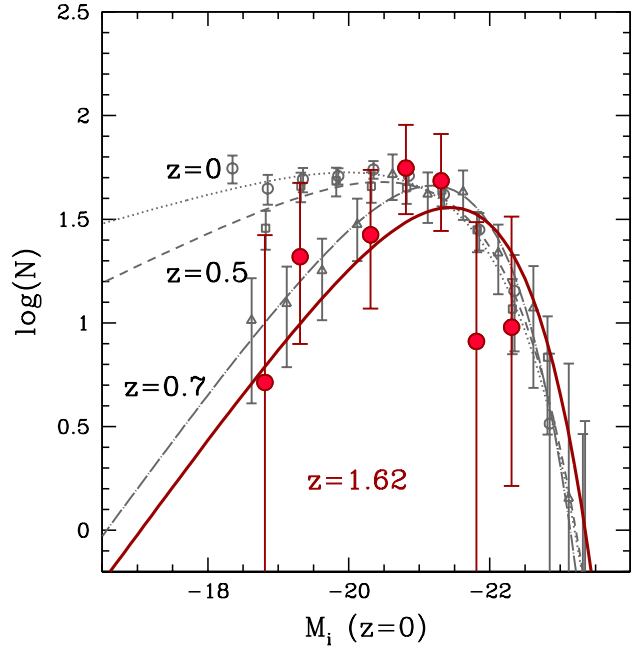}
\caption {The rest-frame $i$-band luminosity function for the cluster
  \clg\ at $z<1.62$ is shown as the solid red circles.  Magnitude bins
  with no objects are not plotted.  The gray open circles, squares,
  and triangle symbols are the LFs for composites of clusters at
  $z<0.06$, $0.4<z<0.6$, and $0.6<z<0.8$ respectively from
  \citet{Rudnick09}.  The error bars are Poisson errors only.  The
  dotted, dash, and dot-dashed lines are the \citet{Schechter76} fits
  to the $z<0.8$ clusters with $\alpha$ left as a free parameter.  The
  solid line is the fit to the $z=1.62$ cluster with $\alpha$ left
  fixed to the $z=0.7$ value.  The magnitudes shown here have all been
  passively evolved to $z=0$ as described in the text.  The LFs
  have also been scaled vertically to have the same integrated
  luminosity.  The measured LF at $z=1.62$ is similar in shape to that
  at $z=0.7$ and shows a significant decline towards fainter
  magnitudes and a lack of bright galaxies.}
\label{Fig:LF}
\end{figure}

In Table~\ref{LF_tab} we give the red sequence LF of \clg\ and show it
in Figure~\ref{Fig:LF} compared to the composite red sequence
luminosity function in clusters at $z=0$, $z=0.5$, and $z=0.7$ from
\citet{Rudnick09}.  As we will discuss shortly, there is clearly
strong evolution in the shape of the LF at $z<0.7$, with little shape
evolution from $z=1.62$ to $z=0.7$.

In \S\ref{Sec:mass_z} we showed that \clg\ is likely the progenitor of
``typical'' clusters in EDisCS and SDSS and so in Figure~\ref{Fig:LF}
we therefore compare the \clg\ LF to the total composites for these
two lower redshift samples.  In Figure~\ref{Fig:LF} we have faded all
the LFs to $z=0$ assuming that the red sequence galaxies all formed at
$z=2.35$ and evolved passively thereafter.  As has been seen before,
the red sequence LF evolves strongly in shape at $z<0.8$ with the
faint galaxy population building up progressively towards lower
redshift
\citep{DeLucia04,Tanaka05,Stott07,DeLucia07,Gilbank08,Rudnick09}.
This has been associated with the late addition of faint galaxies to
the red sequence.

The assumed fading is consistent with the evolution in the mean color
and color scatter of the red sequence in this and other clusters
\citep{Papovich10}.  \citet{Whitaker10} found that passive galaxies at
$z\sim1.6$ have a spread in their ages of about 1~Gyr but this would
result in a spread in the measured fading to $z=0$ of 0.6 mag and
should not affect our results.  The spread in fading is even less if
the galaxies are only faded to $z=0.7$ as we do in \S\ref{Sec:ltot}.

Given this strong evolution at $z<0.8$ it is striking that the
observed LF at $z=1.62$ is remarkably similar in shape to that
$z=0.7$, with the same turnover towards faint magnitudes.  At the same
time, the bright end of the LF at $z=1.62$ appears to be
underpopulated with respect to the individual lower redshift clusters
\citep[see individual cluster LFs from][]{Rudnick09}.  This is similar
to what has been found in a $z=1.46$ cluster \citet{Hilton09}.  Taken
together this implies that galaxies must be added to the bright end
between $z=1.62$ and $z=0.7$ but that the shape of the LF should
remain relatively constant.  In the following section we will discuss
possible scenarios for the evolution of the LF and how it relates to
the evolution of the cluster red sequence.

The \citet{Schechter76} function fits to the EDisCS and SDSS LFs were
presented in \citet{Rudnick09} and were computed with $\alpha$,
\mstar\ and the normalization as free parameters.  Our LF at $z=1.62$
does not have enough signal-to-noise to allow a simultaneous
determination of $\alpha$ and \mstar\ but we attempted to constrain
\mstar\ by fixing $\alpha=0.17$ as determined from the $z=0.7$
clusters.  While the lack of bright galaxies allowed no strong
constraints to be placed on \mstar, the best-fit Schechter function
with a fixed $\alpha=0.17$ (Figure~\ref{Fig:LF}) are statistically
acceptable fits to the observe LF.

\citet{Tanaka10} showed that the red sequence LF of \clg\ as derived
from significantly shallower NIR data appeared similar to that from
groups at $z=1.1$ from \citet{Tanaka08} which in turn had a deficit of
faint galaxies.  At the same time, the very massive $z=1.1$ cluster
from \citet{Tanaka08} had a red sequence LF that appeared similar to
that from SDSS clusters, but was very different from that for \clg.
The strong cluster mass dependence of the LF contrasts to the result
from \citet{Rudnick09} who found only a very weak dependence of red
sequence LF shape on cluster mass.  Indeed, the LF for the most
massive cluster from EDisCS in its high redshift sample
(CL1216.8-1201, $z=0.8$) is consistent with both the $z=0.7$ EDisCS
composite LF and that for \clg\ but inconsistent with the $z=0.5$
EDisCS composite.  As noted in \citet{Tanaka08} and \citet{Rudnick09},
this apparent discrepancy may result from the large cluster-to-cluster
variance in galaxy properties at a fixed mass.  Clearly larger samples
of well-studied clusters are needed at $z>0.5$ over a range of cluster
mass.


\subsection{The Integrated Luminosity of the Red Sequence}
\label{Sec:ltot}

\begin{figure*}
\epsscale{1.15}
\plottwo{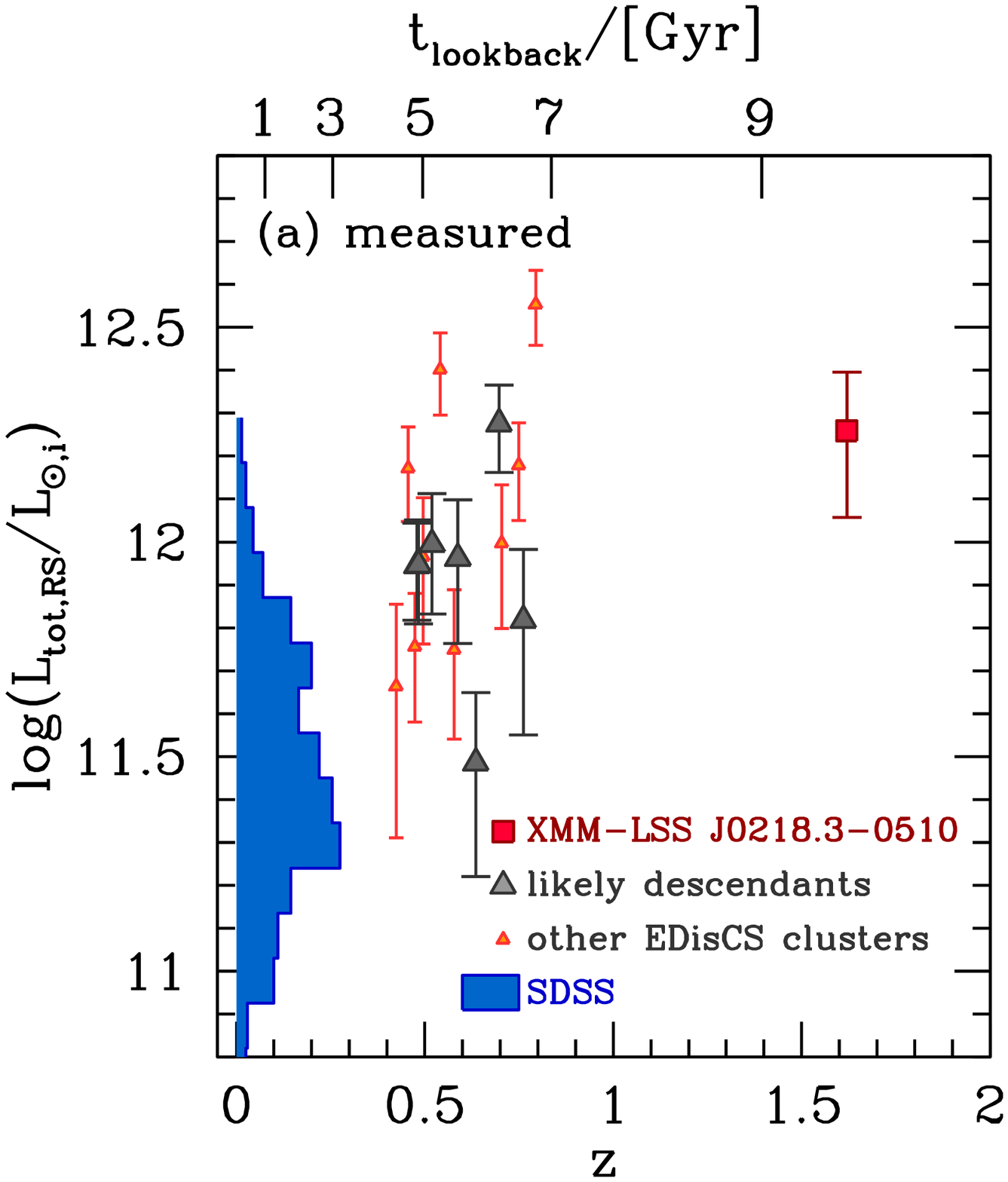}{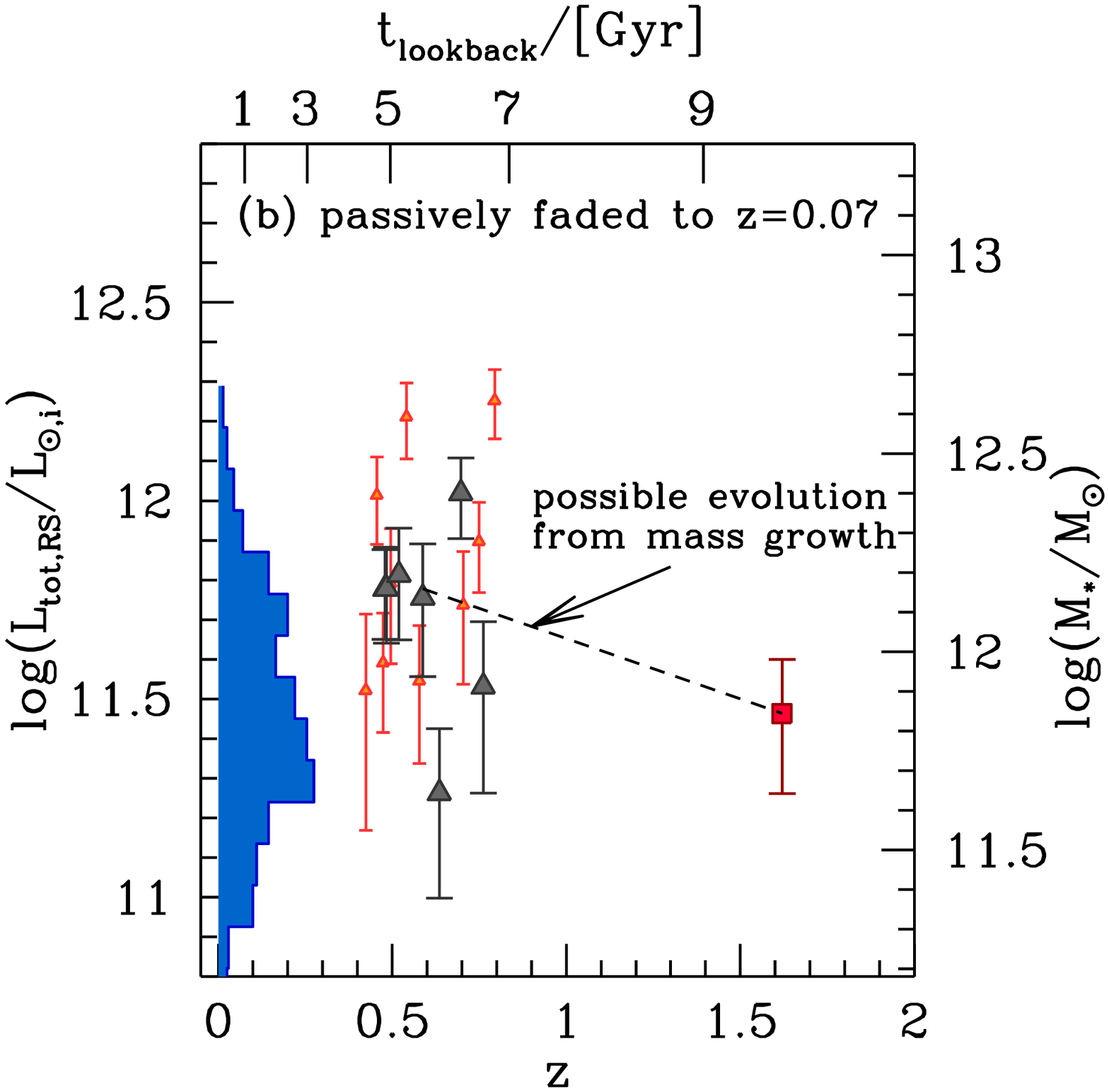}
\caption { The total amount of light on the red sequence, \ltotrs, in
  clusters at $z<1.62$.  The shaded histogram represents the
  distribution total red sequence luminosities from our local SDSS
  sample.  In both panels, the triangles indicate the EDisCS clusters,
  with the larger triangles showing those clusters that are the likely
  intermediate redshift descendants of \clg\ based on the likely mass
  growth from Figure~\ref{Fig:mass_z}.  The errorbars on the EDisCS
  and \clg\ point account for the statistical errors in the individual
  cluster LF determinations.  \textit{Left panel:} \ltotrs, computed
  by integrating the measured red sequence luminosity functions.
  \textit{Right panel:} Both the EDisCS clusters and \clg\ have had
  their \ltotrs\ adjusted by fading the light on the red sequence to
  $z\sim0$ by an amount expected for a simple stellar population with
  $z_{form}=2.35$.  Using this \ml\ correction, we plot the effective
  stellar mass for galaxies at $z=0$ on the y-axis.  The dashed line
  is a least squares fit to the \ltotrs\ values for \clg\ and its
  likely intermediate redshift descendants.
}
\label{Fig:rslight}
\end{figure*}

In addition to determining the evolution (or lack thereof) in the
shape of the luminosity function, a complete description of the growth
of the red sequence also requires a measurement of how the total red
sequence light in clusters increases over time.  To do this we
integrate the observed red sequence LF for \clg\ (\ltotrs) and compare
it to that for clusters at $0.4<z<0.8$ from EDisCS and at $z=0$ from
SDSS.  The results are shown in Figure~\ref{Fig:rslight}.  Our LF
depths at all redshifts have been constructed to have identical faint
limits when corrected for passive luminosity evolution - indeed this
drove our HAWK-I exposure times.  For that reason, integrating the
observed LFs yields an \ltotrs\ measurement that extends to a roughly
constant stellar mass limit.

It is important to compare this cluster with its likely descendants at
low reshift as the most massive clusters also have the red sequences
with the highest \ltotrs\ (see Appendix).  As described in
\S\ref{Sec:mass_z} and as shown in Figure~\ref{Fig:mass_z} the most
likely descendants of \clg\ are typical clusters at $z<1$ with
log$(M/M_\odot)\sim 14.3-15$.  In Figure~\ref{Fig:rslight} we have
highlighted which of the EDisCS clusters are the likely descendants
based on the projected mass evolution.

We compare our clusters within a constant physical aperture of
$r=0.75$~Mpc.  This does not take into account that clusters may
preferentially grow from the inside out \citep[e.g.][]{Balogh00}.  On
the other hand, given the very uncertain mass of \clg\ and the poor
mass determination of some of our SDSS clusters -- due to a small
number of available spectroscopic redshifts -- using a constant metric
aperture is more robust than one which scales as mass, e.g. $R_{200}$.

In the left panel of Figure~\ref{Fig:rslight} it is clear that the
\ltotrs\ of the most luminous cluster red sequences at $z<1.62$ is
comparable to within a few tenths of a dex.  However, we must account
for the expected fading of the stellar populations in the red sequence
galaxies as they age.  To this end, we fade all of the cluster red
sequences to $z=0$ using a simple stellar population formed at
$z=2.35$, which is consistent with the color evolution of the bright
red sequence galaxies over this whole redshift range
\citep{Papovich10}.  The faded total luminosities are shown in the
right panel of Figure~\ref{Fig:rslight}.  Under the assumption that
all the red sequence galaxies at all redshifts have the same SFH, this
roughly converts \ltotrs\ into a stellar mass content on the red
sequence\footnote{\citet{Skelton11} have pointed out that this
  assumption may be inappropriate given the heirarchical build-up of
  red sequence galaxies but we assume it here for simplicity.}.  We
speculate on the evolution of \clg, by performing a least squares fit
to the \ltotrs\ values for \clg\ and the two most massive (and most
luminous) EDisCS clusters.  This is shown as the dashed line in the
right panel of Figure~\ref{Fig:rslight}.

Given the likely evolutionary path, the red sequence in \clg\ grew by
a factor of $\sim 2$ in light or stellar mass during the $\approx
4$Gyr between $z=1.62$ and $z=0.6$.  In \S\ref{Sec:discuss} we will
discuss how to reconcile this rapid growth in the luminosity with the
lack of shape evolution between $z=1.62$ and $z=0.7$.  Assuming that
the evolutionary path in Figure~\ref{Fig:rslight} continues to $z=0$
then the cluster will grow by an addition 50\% between $z=0.6$ and
$z=0$ or a factor of three in total since $z=1.62$.

As a cautionary note, given the rapid evolution in the shape of the LF
at $z<0.6$ it is likely incorrect to assume a constant amount of
fading for all galaxies, as those added more recently to the red
sequence will fade more rapidly with time.  To approximate this effect
we varied the amount of fading assuming a $z_{form}$ for \clg\ ranging
from $z=2$ to 3 and for the EDisCS clusters ranging from $z=1$ to 2.
While the exact trend of \ltotrs\ with $z$ depends on the exact mix of
$z_{form}$, the change in maximum growth in \ltotrs\ was $\sim 0.1$
dex over the full redshift range.

\section{Discussion}
\label{Sec:discuss}

Our first result, that there is a deficit of faint red galaxies in
\clg\ may result naturally from a scenario in which galaxy star
formation is quenched once a galaxy's total mass moves above $10^{12}
M_\odot$ and it forms a hot gas halo.  This naturally predicts that
the most massive galaxies are quenched first and that there should be
relatively few low-mass passive galaxies at high redshift
\citep{Gabor12}.

In explaining our other findings on the growth of the red sequence
over time we must explain some apparently contradictory results.
First, we find that the LF of red sequence galaxies in clusters
evolves very little \textit{in shape} from $z=1.62$ to $z=0.6$.  At
redshifts lower than this, however, the shape evolves rapidly, such
that the faint end slope becomes shallower, eventually matching the
$z=0$ value (Figure~\ref{Fig:LF}).  At the same time,
Figure~\ref{Fig:rslight} alone implies that the total stellar mass on
the red sequence appears to increase by a factor of $\sim 2$ during
the $\sim 4$~Gyr from $z=1.62$ to $z=0.6$ and then grows by only 50\%
over the remaining $\sim 6$~Gyr to the present day.  In
\citet{Rudnick09} it was shown that adding the minimal possible number
of galaxies to the red sequence at $z<0.8$ would actually cause the
predicted \ltotrs\ in SDSS clusters to be too high.  Resolving this
discrepancy can be accomplished by assuming that a significant
fraction of the stars in galaxies that are added to the red sequence
at $z<0.8$ end up as intracluster stars.  At $z>0.7$ we appear to have
the opposite problem.  The total light in the clusters must grow
rapidly, but without changing the shape.

\begin{figure*}
\plotone{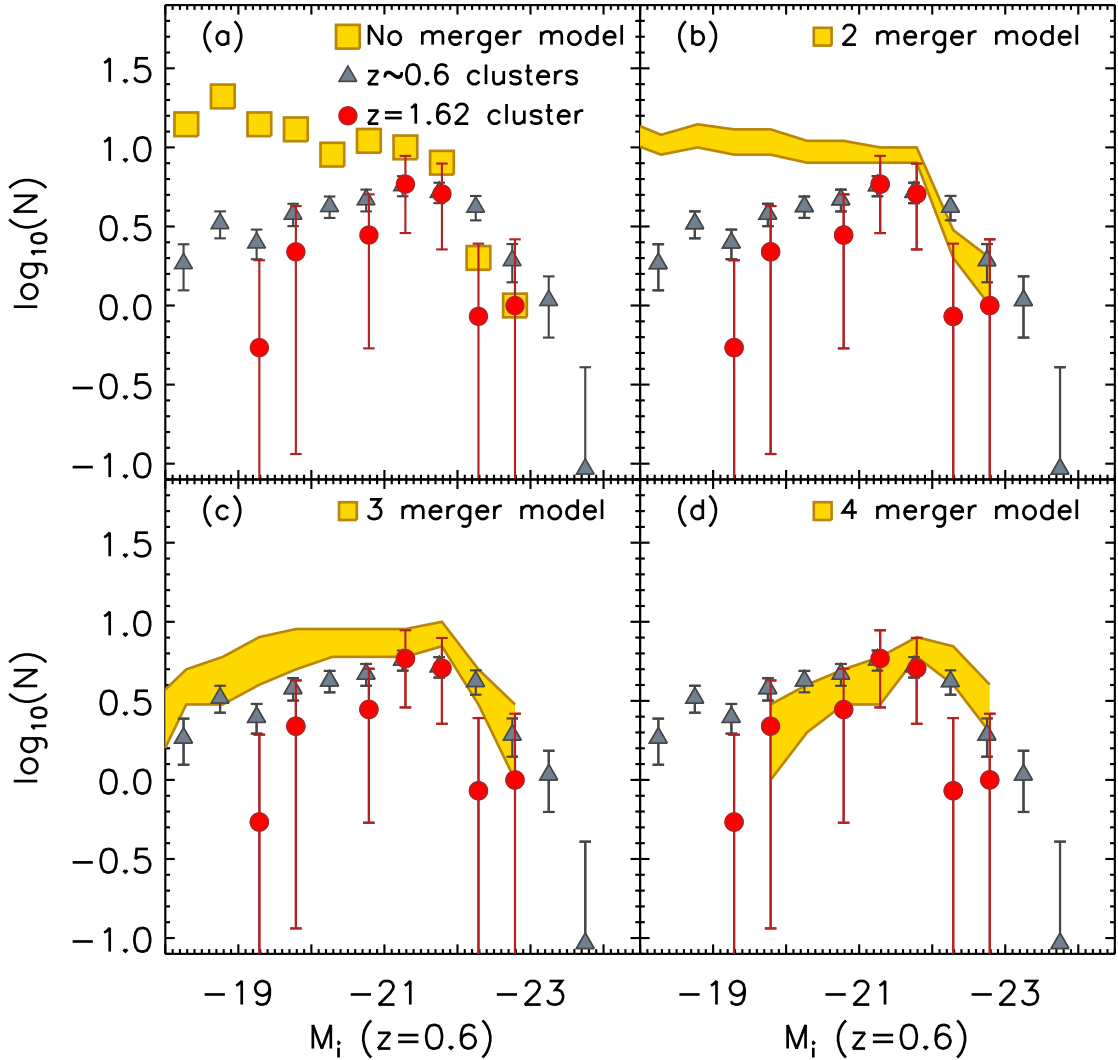}
\caption {A demonstration of how merging and accretion will affect the
  evolution in the red sequence LF between $z=1.62$ and $z=0.6$.  In
  all panels the vertical scaling of the luminosity functions are
  accurate in an absolute sense.  The x-axis refers to the magnitude
  that galaxies are predicted to have at $z=0.6$ assuming the SFHs
  given in the text.  The red circles show the LF of \clg.  The gray
  triangles show the composite EDisCS red sequence LF scaled in
  luminosity to the mean \ltotrs\ of the likely descendant clusters at
  $z~0.6$.  These points are the same in each panel.  The panels show
  a set of model predictions for the LF at $z=0.6$ that assume that
  the cluster accretes enough recently quenched galaxies to account
  for the evolution in total red sequence light at $M_i(z=0.6)<-19$.
  The squares in panel (a) show the predicted LF in the absence of
  merging of passive galaxies.  The filled bands in panels (b), (c),
  and (d), represent how the predicted LF will change if every passive
  galaxy randomly merges two, three, or four times between $z=1.62$
  and $z=0.6$.  The extent of the bands corresponds to the 25 and 75\%
  confidence intervals on the predicted LF resulting from a monte
  carlo simulation as described in the text.  The extent of the bands
  in magnitude correspond to where the simulated LFs contain a median
  of 1 galaxy.  The cluster LF shape and normalization at $z=0.6$ can
  be approximately reproduced if every galaxy merges approximately
  three to four times over the intervening $\sim 4$ Gyr between
  $z=1.62$ and $z=0.6$.}
\label{Fig:lf_sim}
\end{figure*}

At first glance, resolving this problem is difficult.  There are no
blue galaxies in the field at this redshift luminous enough to fade
onto the bright end of the red sequence in this cluster
\citep{Papovich10}, which implies that most of the mass must be added
in the form of fainter galaxies.  However, this would result in a
shape change of the LF that is not seen between $z=1.62$ and $z=0.7$.
A hint perhaps lies in the lack of red galaxies in \clg\ brighter than
$M_i=-24.2$, or $M_{i,fade}=-22.25$, as has also been seen in
\citet{Hilton09}.  This is in contrast to the individual EDisCS
clusters, which are populated with galaxies to at least 0.5 magnitudes
brighter than this, when accounting for the passive fading of the
stellar populations \citep[See individual LFs from][]{Rudnick09}.
Therefore, despite the lack of blue luminous galaxies that could fade
onto the red sequence, the bright red sequence population must grow
between $z=1.62$ and 0.7.  One solution to this problem is for
galaxies in the blue cloud at fainter magnitudes ($\sim$\lstar) to
rapidly form stars and then be quenched, all on short enough
timescales to prevent them from showing up in the observed population.
The amount of stars that need to be formed in such an episode,
however, would require SFRs much in excess of what is measured for
galaxies in this field at $z\sim 1.6$ from MIPS 24\micron\ data
\citep{Tran10}.  

\subsection{The importance of mergers}

A more likely explanation is that the fainter blue galaxies fall into
the cluster, have their star formation suppressed somwhere during the
process (in groups or the cluster itself), and migrate to the faint
end of the red sequence, where they subsequently merge with other red
sequence galaxies and increase their mass.  It is reasonable that the
SFRs of infalling galaxies are suppressed as they fall into the
cluster environment, via ram-pressure stripping \citep{Gunn72} or
galaxy strangulation \citep{Larson80}.  Indeed \citet{Pierre11} have
found that \clg\ has a diffuse x-ray component indicating the presence
of a heated (but possibly underluminous) IGM.  Such a scenario would
have the effect of offsetting the addition of galaxies to the red
sequence at faint magnitudes by merging them up the red sequence and
hence preserving the shape of the LF, while at the same time allowing
the cluster to grow its total red sequence stellar mass to agree with
the most massive EDisCS clusters.

To test this we perform a simulation where we add galaxies to the red
sequence between $z=1.62$ and $z=0.6$ and then test how the resultant
red sequence LF shape and normalization depends on the different
number of mergers per galaxy.  Because the shape and normlization are
important for this test, we compare the LF for \clg\ to the composite
EDisCS LF normalized to the mean \ltotrs\ of the likely descendant
clusters.

We assume that infalling field galaxies have a constant SFR until a
time $t_{cut}$ before they are rapidly quenched.  This quenching
happens a time $t_{delay}$ before they are added to the cluster.  This
insures that they are red by the time they enter the cluster.  During
the quenching process they lose $A_i$ magnitudes of internal
extinction.  For the purpose of this test we choose $t_{cut}=3$~Gyr,
$t_{delay}=1.0$~Gyr, and $A_i=0.7$ mag (corresponding to $A_V=1$ for a
\citet{Calzetti00} attenuation curve), although we note that our
results are not sensitive to the exact values adopted (see below).
This scenario assumes that the SFRs of galaxies are suppressed before
falling into the cluster environment, and that in the transformation
to red sequence galaxies they lose some amount of dust extinction.
This general scenario is consistent with previous works
\citep[e.g.][]{Poggianti06,DeLucia07,Rudnick09,DeLucia07,McGee11} and
results in galaxies with colors on the red sequence by $z=0.6$.

Over the 3.8~Gyr between $z=1.62$ and 0.6 the cluster red sequence of
\clg\ has to increase its total stellar mass by a factor of 2 to match
\ltotrs\ for the likely descendants at $z\sim 0.6$.  We split the
necessary mass increase equally between 3 identical intervals of time
over this redshift range.  In each interval we draw galaxies from the
evolving blue field-galaxy LF from \citet{Salimbeni08} and evolve them
using the SFH parameters above to predict their luminosities at
$z=0.6$.  The Salimbeni LF is computed in the rest-frame B-band.  We
convert it to the rest-frame $i$-band assuming a constant SFH with 1
mag of extinction.  The result of this process is shown in
Figure~\ref{Fig:lf_sim}a as the yellow squares, which are compared to
the \clg\ and EDisCS LF.  There we can see that such a scenario of
simply accreting and quenching galaxies from the star-forming field
population to make up the observed total luminosity increase would
produce a faint end slope much steeper than what is observed in
clusters at $z=0.6$.

We then merge each galaxy with a random second galaxy and perform this
up to four times.  We perform 100 monte-carlo iterations of these
mergers and show the 25 and 75\% limits of the distribution of the
resultant LF as the yellow bands in Figure~\ref{Fig:lf_sim}b, c, and d
for two, three, and four mergers per galaxy respectively.  It appears
that merging each red sequence galaxy three to four times with another
red sequence galaxy results in an LF that is similar to that observed
at $z=0.6$.  This implies a merger rate of $\sim 1$ per Gyr over this
time period, integrated over all merger mass ratios.  The faint-end
slope for this model has the necessary turnover but the bright end is
still underpopulated.  If galaxies only merge twice then the faint-end
slope is too steep.  If galaxies merge four times, then the faint end
turns over a little bit too rapidly but the bright end is somewhat
above the observations yet does not extend to the brightest magnitudes
seen in EDisCS.  

These remaining challenges may imply that our model is too simplistic
in its treatement of the mass ratios of mergers.  For example, in a
perfect world the mass ratios should be drawn from the LFs of
infalling galaxies and not from the general field population.  Also,
\citet{DeLucia12} find that major mergers of massive galaxies in
clusters made be preferred based on their results using N-body
simulations with a semi-analytic galaxy formation model.  An additonal
potentially important unmodeled ingredient could the inclusion of the
rapid buildup of stellar mass in massive cluster galaxies that are
undergoing significant star formation but which must cease their star
formation soon after we observe them \citep{Tran10}.  Since the number
of dusty star-forming galaxies increases rapidly towards higher
redshift, both in the field \citep[e.g.][]{lefloch05} and in clusters
\citep{Saintonge08,Finn10}, their rapid quenching may be a plausible
mechanism to build up the bright end of the red sequence luminosity
function.  

It is worth noting that the exact number of mergers needed to match
the LF is affected greatly by the slope of the faint-end of the blue
field LF.  If we assume $\alpha=-1.0$ instead of -1.39 as in
\citet{Salimbeni08} then the required number of mergers goes down to
1--2.  The results of this model are also dependent on the faint limit
from which we draw field galaxies.  We use a faint limit of -14 but
note that going to -16 would require two mergers to sufficiently
deplete the faint-end of the LF.  The results are less dependent on
the SFH parameters $t_{cut}, t_{delay}$, and $A_i$.  For example
$t_{cut}<2$~Gyr and $A_i<0.5$ both tend to favor four mergers, with
three being ruled out.\footnote{$A_i=0.7$ corresponds to the typical opacity
of galactic disks.}.

The importance of merging in growing the red sequence is consistent
with \citet{Papovich11} and \citet{Lotz11} who study this cluster with
WFC3 imaging from the CANDELS program.  \citet{Lotz11} found that
\clg\ has a $\sim 10$ times higher merger rate than analogously
selected galaxies in the coeval field, and that the close pairs are
dominated by galaxies with a high mass ratio, i.e. minor mergers.  The
merger rate cited by \citet{Lotz11} is 2 per Gyr, which is consistent
with our estimate of 1 per Gyr given the considerable uncertainties in
both estimates, e.g. in the dependence of our derived merger rate on
the field luminosity function.  In addition \citet{Papovich11} found
using the same CANDELS data that the massive red sequence galaxies in
\clg\ require extensive dissipationless, and perhaps minor, mergers in
order to simultaneously match the stellar mass, size, ellipticity, and
color of cluster red sequence galaxies at $z<1$.

There is some direct evidence for merging of red sequence galaxies in
a merging cluster system at $z=0.8$ \citep{vandokkum99,Tran05}.
Likewise, \citet{white07} found that to evolve the Halo Occupation
Distribution for red luminous satellite galaxies (i.e. cluster red
sequence galaxies) from $z=0.9$ to 0.5 requires that 1/3 of these
galaxies must merge or undergo disruption in massive halos.  This is
again broadly consistent with the scenario that we propose here as the
\citet{white07} study includes redshifts where the growth in the red
sequence has become more gradual.  While merging is unlikely in a
relaxed massive cluster due to the high velocities, \clg\ will be
rapidly growing with time, accreting other clusters and groups along
the way.  It is not relaxed, which implies that the galaxy-galaxy
velocities may be lower than expected given the cluster mass.  The
relative velocities in the infalling groups will also be lower than in
the final cluster and at group scales the merging cross-section may be
quite high.  Indeed \citet{Tran08} find that the massive red galaxies
in a set of merging groups at $z\sim 0.37$ are themselves experiencing
dissipationless mergers that will significantly grow their mass.  By
$z=0.8$ the most massive clusters in EDisCS have high velocity
dispersions \citep{Milvang08}, detected intracluster light
\citep{Guennou11}, and strong gravitational lensing \citep{White05},
implying that the cross section for merging may be much lower.

\subsection{Additional Caveats}

Our main uncertainty results from this study being based on one high
redshift cluster.  As pointed out in \S\ref{SubSec:LF} there are
differences in the LF of massive clusters at $0.8<z<1.1$ and we can
expect that these differences might be more extreme in the
less-developed cluster population at $z>1.5$.  Obviously, having larger
samples of high redshift clusters with deep NIR data will be crucial
if we wish to obtain a representative picture of galaxy evolution in
clusters at these epochs.

An additional uncertainty in our result regards the identification of
red sequence galaxies with passively evolving ones.  At $z=0.8$ this
is true at the $\sim 85\%$ level (Rudnick et al. in prep.).  At
$z=1.62$, \citet{Tran10} found that some of the red sequence galaxies
were 24\micron\ emitters, indicating that a small number of red
galaxies could be dusty star formers.  \citet{Papovich11} and
\citet{Quadri12} used combinations of rest-frame optical/NIR colors to
separate red galaxies that are quiescient from those that are obscured
star-formers \cite[see also][]{Williams09}) and found that the
contamination on the red sequence from dust-obscured objects is
$\sim20\%$ for bright red galaxies.  Given the similar fractions of
obscured red galaxies at both redshifts we conclude that this will not
cause a significant error in our conclusions.

The mass estimate for our cluster is also uncertain, with a range
between measurements of $M\sim 7.7\times 10^{13}-1\times
10^{14}$\msol.  We adopted to lower of these two - from the x-ray
detection - for the analysis in this paper.  If we instead chose the
mean of these two masses, then the projected mass growth tracks
(yellow band in Figure~\ref{Fig:mass_z}) would shift upward but remain
roughly parallel to the original.  The likely descendant clusters at
$z\sim 0.6$ would then be the most massive clusters at every epoch
with $M>10^{15}M_\odot$ although in reality such extreme descendants
are unlikely due to the steepness of the cluster mass function.  The
EDisCS LF shape is not dependent on cluster mass \citep{Rudnick09},
and so the observed lack of evolution in the shape should remain
unchanged regardless of the expected descendants.  If we adopt the
mean mass then the expected luminosity evolution in \S\ref{Sec:ltot}
and Figure~\ref{Fig:rslight} would be a factor 6 increase to $z\sim 0.6$
and almost no luminosity evolution to $z=0$.  As this is very extreme
we are further confident that the lower mass estimate is more
appropriate.

A final caveat is our reliance on statistical background subtraction
to determine membership.  These should have equal or greater
reliability when compared to photometric redshifts
(\S\ref{Sec:member}) but spectroscopic redshifts are indeed sparse and
will help to improve the constraints on the LF, especially at the
bright end where continuum redshifts are feasible.  Ample spectroscopy
will also allow us to better measure the velocity dispersion and
dynamical state of the cluster.  Likewise, deep medium-band NIR
imaging with 6 or 8-meter telescopes, can provide very precise
photometric redshifts that would improve our analysis
\citep{Whitaker11}.

\section{Summary and Conclusion}
\label{Sec:summ}

We have presented the rest-frame $i$-band luminosity function of red
sequence galaxies of \clg, a cluster at $z=1.62$, as measured using
deep HAWK-I observations from the VLT and a $K_s$ selected catalog.
At this redshift $Y-J$ straddles the Balmer/4000\AA\ break and $K_s$
is near in wavelength to rest-frame $i$.  Our conclusions are as
follows:

\begin{itemize}

\item \clg\ has a strong red sequence at bright magnitudes.  Starting
  well brighter than our 90\% completeness limit we find a lack of
  faint red galaxies.  This conclusion is not dependent on our
  membership identification scheme as these galaxies simply do not
  exist in this area of the sky.  The cluster red sequence hosts the
  objects with the reddest observed colors in this field and there are
  very few at faint magnitudes.

\item We derive a luminosity function for the red sequence cluster
  members and compare it to analogously constructed LFs for clusters
  at $z=0$ from SDSS and $0.4<z<0.8$ from EDisCS.  When corrected for
  passive evolution in the luminosities, we find that the shape of the
  \clg\ LF is indistinguishable from the cluster red sequence LF at
  $z=0.7$ and exhibits the same turnover to faint magnitudes that has
  been noted by other authors in $z<1$ clusters.  At $z<0.7$ the faint
  end of the luminosity function starts to rise and has a near flat
  slope at $z=0$.  We also find that \clg\ has a lack of luminous red
  galaxies when compared to clusters from EDisCS.

\item The integral of the measured red sequence LF shows that
  \ltotrs\ for \clg\ is as high as the most luminous EDisCS cluster at
  $0.4<z<0.8$ and as the most luminous clusters in SDSS.  However,
  when the expected evolution in the the stellar mass-to-light ratio
  is accounted for, \clg\ has at least 3 times less stellar mass on
  the red sequence compared to its likely descendant clusters in SDSS
  and a factor of 2 less stellar mass than its likely descendants in
  EDisCS.

\item We attempt to explain the large growth in the total luminosity
  between $z=1.62$ and $z=0.6$ while simultaneously preventing the
  buildup of faint red galaxies and also growing massive red ones.  We
  test a simple model in which the cluster accretes galaxies from the
  blue field population that are quenched and then merge on the red
  sequence.  If every cluster red sequence galaxy mergers $\sim 3-4$
  times with another red sequence galaxy, we can satisfy all of the
  above constraints.  This corresponds to a merger rate of $\sim
  1$~Gyr$^{-1}$ integrated over all mass ratios.  This also agrees
  with independent merger rates from this cluster.

\end{itemize}

There are some limitations to our analysis.  First, our membership
information is limited due to the difficulty in obtaining
spectroscopic membership information.  NIR spectrographs on 6 and
8-meter telescopes will be crucial here.  We also only have one
cluster with an LF of moderate signal-to-noise.  Obviously, increasing
the sample of clusters will allow for a more precise determination of
the mean cluster LF at these redshifts and will let us determine if
\clg\ is typical of the high redshift cluster population.

\acknowledgements

GHR would like to thank Hans-Walter Rix and the Max-Planck-Institute
for Astronomy in Heidelberg for its warm hospitality during work on
parts of this paper.  This material is based upon work supported by
the National Science Foundation under Award No. EPS-0903806 and
matching support from the State of Kansas through Kansas Technology
Enterprise Corporation.  The authors thank Chris Lidman for useful
discussions regarding the processing of HAWK-I data.

\clearpage


\clearpage
\appendix

\section{The Use of the Integrated Red Sequence Light as an Estimator of the Dynamical Mass.}
\label{Sec:app}

Since the dynamical state of our cluster is likely not relaxed, we
seek to estimate the mass using a different method.  In
Figure~\ref{Fig:mass} we show the correlation of \ltotrs\ in a
$r=0.75$~Mpc aperture vs. weak lensing mass \citep{Clowe06} and mass
derived from the velocity disperion \citep{Milvang08} for the EDisCS
clusters at $0.4<z<0.8$.  There is a clear correlation between lensing
mass and \ltotrs\ but a significantly poorer relation with mass
derived from the velocity disperion, even though this dispersion is
measured from 30--50 members.  The horizontal line shows the value for
\clg, where we have corrected the luminosities for passive evolution
to $z=0.6$.  If we assume the same ratio of dark matter mass to
stellar mass on the red sequence at $z=0.6$ and $z=1.62$ then it would
appear that \clg\ has $M\sim 2.5\times 10^{14}M_\odot$, which is in
excellent agreement with the value of $M\sim 1-4\times 10^{14}M_\odot$
from \citet{Papovich10} and \citet{Pierre11} as determined from the
x-ray luminosity on the low side and the very uncertain velocity
dispersion ($\sigma=860\pm490$km~s$^{-1}$) on this high side.  This
opens the possibily of using \ltotrs\ as a proxy for mass in clusters
that we are now selecting with this method, but which do not have
spectroscopy sufficient to determine a velocity dispersion.

\renewcommand\thefigure{A\arabic{figure}}
\setcounter{figure}{0}
\begin{figure}
\epsscale{0.6}
\plotone{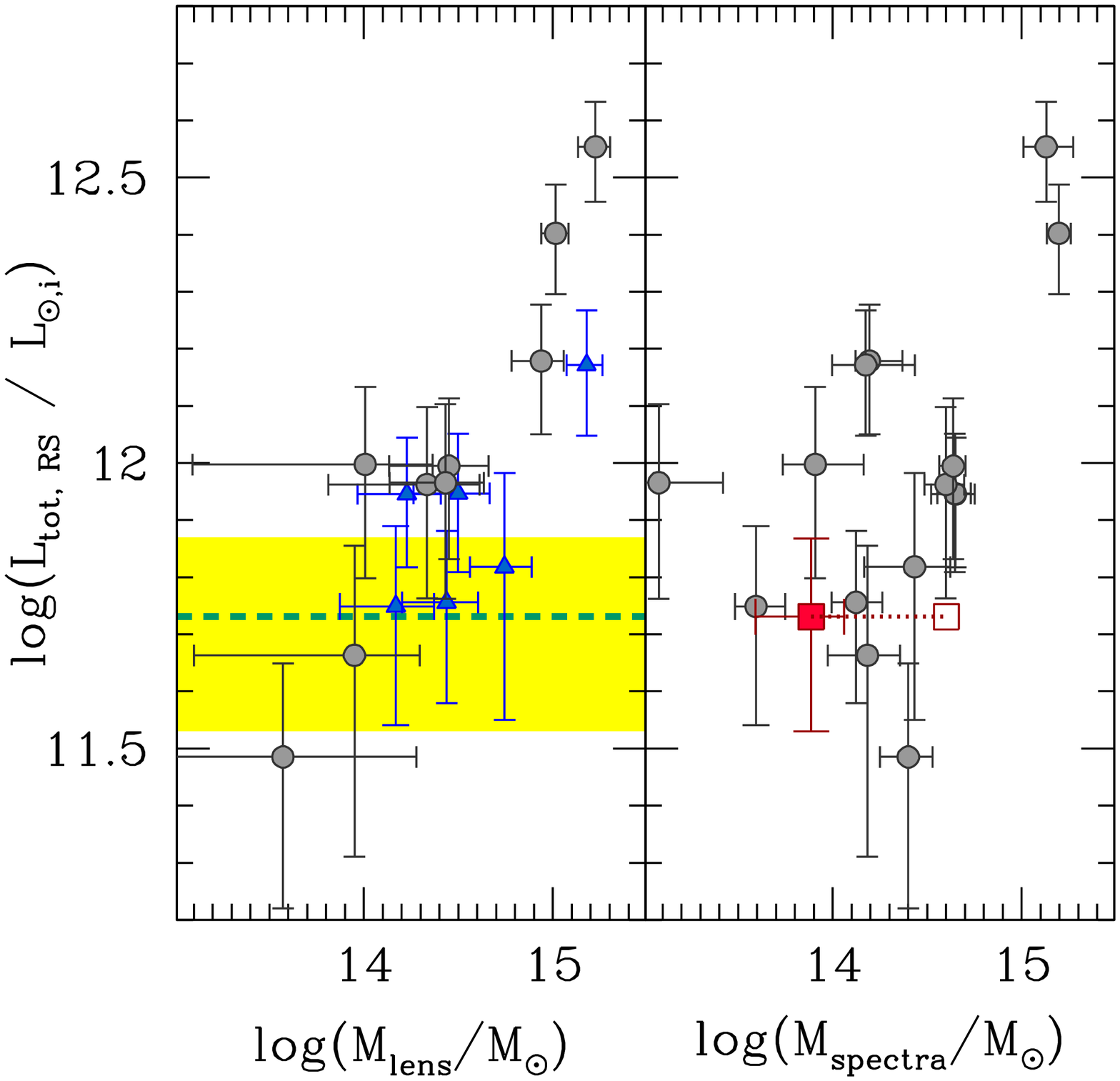}
\caption {A comparison of the dynamical and weak-lensing mass with the
  integral of the red sequence luminosity \ltotrs.  The circle and
  triangle points are values for the EDisCS clusters $0.4<z<0.8$.
  \textit{Left panel:} The EDisCS masses are derived from weak lensing
  estimates \citep{Clowe06}, where the blue points are those where
  substructures may contaminate the lensing signal.  The horizontal
  dashed line and yellow band (left panel) and red point (right panel)
  gives \ltotrs\ and its uncertainty for \clg, where the luminosities
  have been passively faded to $z=0.6$.  \textit{Right panel:} The
  EDisCS masses in this panel have been derived from the velocity
  dispersion using 30-50 members per system from \citet{Milvang08}.
  The red solid square represents the x-ray derived mass for
  \clg\ \citep{Pierre11}.  The open square connected to the solid by a
  dotted line indicates the dynamical mass estimate from
  \citet{Papovich10}.  The disagreement between the two likely
  reflects the unrelaxed dynamical state of the system.  All
  luminosities are computed in a circular aperture with a radius of
  0.75 physical Mpc.  The EDisCS luminosities have all had a small
  passive evolution correction to $z=0.6$.  The correlation of
  \ltotrs\ and cluster mass is significant for lensing mass, but less
  so for mass from the velocity dispersions.}
\label{Fig:mass}
\end{figure}

\end{document}